\documentstyle [12pt]{article}
\newcommand{\etal}{{et al.}~}

\newcommand{\p}{\partial}
\newcommand{\f}{\frac}

\newcommand{\Om}{\Omega}
\newcommand{\de}{\delta}
\newcommand{\eps}{\epsilon}

\newcommand{\al}{\alpha}

\newcommand{\fPhi}{\tilde{\Phi}}
\newcommand{\fcalP}{\tilde{\cal P}}
\newcommand{\calP}{{\cal P}}

\newcommand{\bfx}{{\bf x}}

\newcommand{\bfv}{{\bf v}}
\newcommand{\bff}{{\bf f}}
\newcommand{\bfq}{{\bf q}}
\newcommand{\bfp}{{\bf p}}
\newcommand{\bfg}{{\bf g}}
\newcommand{\bfA}{{\bf A}}
\newcommand{\bfB}{{\bf B}}
\newcommand{\bfR}{{\bf R}}
\newcommand{\bfS}{{\bf S}}
\newcommand{\bfT}{{\bf T}}
\newcommand{\bfu}{{\bf u}}
\newcommand{\bfr}{{\bf r}}
\newcommand{\calS}{{\cal S}}
\newcommand{\bc}{\begin{center}}
\newcommand{\be}{\begin{equation}}
\newcommand{\ee}{\end{equation}}
\newcommand{\ec}{\end{center}}

\title{{\LARGE {\bf LAGRANGIAN DYNAMICS IN NON--FLAT UNIVERSES
AND NON--LINEAR GRAVITATIONAL EVOLUTION}}\\}
\vspace{1cm}
\author{ {\bf Paolo CATELAN} \\ \\
{\it Department of Physics--Astrophysics, Nuclear Physics Laboratory}\\
{\it Keble Road, Oxford OX1 3RH, UK} \\ }
\begin{document}
\maketitle
\vspace{1.5cm}
\bc
{\it MNRAS}, in press
\ec
\vspace{1.5cm}
\bc
{\it OUAST/94/19}
\ec
\newpage
\vspace{1cm}
\bc
\section*{Abstract}
\ec
We present a new formalism which allows to
derive the general Lagrangian dynamical equations for the motion of gravitating
 particles in a non--flat Friedmann universe with arbitrary density parameter
$\Om$ and no cosmological constant. We treat the set of particles as a
Newtonian
collisionless fluid. The non--linear dynamical evolution of the fluid
element trajectories is then described up to the third--order expansion in
Lagrangian coordinates. This work generalizes recent investigations carried out
by Bouchet \etal (1992) and Buchert (1994).

\vspace{0.5cm}
\noindent{\em Subject headings:} Galaxies: clustering -- large--scale
structure of the Universe
\vspace{1.0cm}

\section{Introduction}
The most widely accepted hypothesis about the formation of large--scale
structures is that galaxies and clusters of galaxies formed by gravitational
collapse around primordial slightly overdense mass fluctuations in the
Universe. One way to link these initial conditions to the final mass
distribution is to attempt a solution of the equations for
the matter fluctuation field $\de$ and the peculiar velocity field $\bfv$,
namely the Euler equation, the continuity equation and the Poisson equation.

Linear theory easily provides solutions even for cosmological models with
arbitrary density parameter $\Om$ (e.g. Peebles 1980), but when
the perturbation amplitudes approach unity, non--linear gravitational
effects become very important. It is undoubtedly impossible to
follow the  non--linear stage of the
gravitational collapse  in an exact analytical
way, and one is forced to use approximation techniques such as
perturbation theory where, for instance, the systematic expansion of the
density solution is obtained by writing $\de=\sum\de^{(n)}$, where
$\de^{(n)}=O(\de^{(1)n})$, $\de^{(1)}$ corresponding to the linear solution
(Fry 1984; Goroff \etal 1986; Juszkiewicz, Bouchet \& Colombi 1993;
Catelan \& Moscardini 1994a,b; Bernardeau 1992, 1994; Catelan et al. 1995).

Usually, these analyses are intrinsically $Eulerian$, the fundamental
quantities being the density and velocity fields evaluated at the (comoving)
Eulerian coordinate $\bfx$; also, with the exception of Bernardeau (1992, 1994)
and Catelan et al. (1995), they have been confined within the limits
of the Einstein--de Sitter cosmology, mainly because the condition $\Omega=1$
enormously eases the investigations of solutions of the dynamical equations for
$\de$ and $\bfv$: apart from theoretical expectations, there is however
no definitive observational evidence that our Universe is really flat
(see e.g. Peebles 1991; Coles \& Ellis 1994).

It has been recognized that the problem of giving an analytical description
of the non--linear process of gravitational clustering simplifies when
formulated in terms of $Lagrangian$ coordinates rather than the standard
Eulerian ones: Zel'dovich (1970a, b) first proposed to approximately
describe the weakly non--linear regime of density evolution in terms of the
departure from to the Lagrangian (initial) positions of the fluid
elements. The Zel'dovich approximation is now widely used in
cosmology, showing also to be extremely useful in reconstruction methods
of initial conditions from velocity data (e.g. Nusser \& Dekel 1992).
However, only recently it has been fully understood that the whole
dynamics of gravitational clustering may be suitably described in terms of
the displacement field $\bfS$, which turns out, in the Lagrangian approach,
to be the only underlying fundamental field. Buchert (1989, 1992)
indeed derives the exact equations governing the evolution of the
displacement $\bfS$ (therefore of the density and velocity fields): since, as
in the Eulerian case, it is impossible to work out the general solution $\bfS$,
a perturbative approach is again introduced. The key novelty with respect
to the Eulerian approach is that one searches for solutions of {\it perturbed
trajectories} about the linear (initial) displacement $\bfS^{(1)}\,$:
$\,\bfS = \sum\bfS^{(n)}$ where $\bfS^{(n)} = o(\bfS^{(1)})$ (see
Moutarde \etal 1991).
The important point is that a $slight$ perturbation of the Lagrangian
particle paths carries a large amount of non--linear information
about the corresponding Eulerian evolved observables, since the
Lagrangian picture is intrinsically non--linear in the density field.

Solutions up to the third--order Lagrangian approximation have been
obtained (the first--order solution corresponding
to the Zel'dovich approximation), although limited to the case of an
Einstein--de Sitter model (Buchert \& Ehlers 1993; Buchert 1994).
The higher accuracy of Lagrangian perturbative methods as compared to
other currently studied approximation ansatzs [such as the frozen flow
(Matarrese \etal 1992) and the linear potential approximations
(Brainerd, Scherrer \& Villumsen 1993; Bagla \& Padmanabhan 1994)] is
discussed in Munshi \& Starobinsky (1994), Bernardeau \etal (1994) and
Munshi, Sahni \& Starobinsky (1994), again in the framework of an
Einstein--de Sitter cosmology. Comparisons with $N$--body simulations
in the fully developed non--linear clustering are displayed in
Moutarde \etal (1991), Coles, Melott \& Shandarin (1993) and
Melott \etal (1994).

The second--order Lagrangian solution for generic non--flat Friedmann
models
has been derived by Bouchet \etal (1992; hereafter BJCP), where particular
emphasis on the connection with the Eulerian formulation has been given.
Further attempts to extend the Lagrangian formalism to models with arbitrary
density parameter may be found in Gramann (1993) and Lachi\`eze--Rey (1993b),
which, however, lead to not completely correct conclusions.
Matarrese, Pantano \& Saez (1994a,b), developed a relativistic Lagrangian
treatment of the non--linear dynamics of an irrotational
collisionless fluid, which reduced to the standard Newtonian
approach on sub--horizon scales, but is also suitable for the description of
perturbations on super--horizon scales.

In this work, we present a new $Lagrangian\,$ formalism
which enables one to easily
derive the exact dynamical equations governing a pressureless
Newtonian gravitating fluid in an expanding universe with arbitrary
density parameter $\Omega$ and no cosmological constant.
We then describe the non--linear evolution of perturbations
up to the third--order Lagrangian approximation. The first--order
solution corresponds, of course, to the Zel'dovich approximation, and
the second--order one to the BJCP approximation. The third--order
solution is then derived in detail for arbitrary initial conditions.
It consists of three independent (growing) modes, two being purely longitudinal
and one purely transversal,
in such a way that a fluid which is irrotational in Eulerian space is not
so in the corresponding Lagrangian space. This problem
was first addressed in Buchert (1994),
where the consequences of
requiring irrotationality of the fluid motion in Lagrangian space on the
initial conditions are explored in detail.

The layout of this paper is as follows. In section 2 the Lagrangian
approach is reviewed. In section 3, the general Lagrangian equations
describing the dynamical evolution of a collisionless Newtonian
self--gravitating fluid in an expanding universe with arbitrary
density parameter $\Om$ are obtained. In section 4, after
rederiving in our formalism the Zel'dovich and the BJCP approximations,
we work out the third--order Lagrangian approximation. Our conclusions
are presented in section 5. To avoid an excess of mathematical
contortions in the text, four technical appendices are given.
\section{Lagrangian Formulation}
Let us consider a Newtonian pressureless self--gravitating fluid
embedded in an expanding universe with arbitrary density parameter $\Om$.
We assume that the cosmological constant is exactly zero. We have in mind
that such a fluid mimics the behavior of matter on scales smaller
than the horizon scale and that, around its primordial density perturbations
$\de$, the present luminous
objects like galaxies or clusters of galaxies started to grow according
to a gravitational instability  process. We indicate by $\bfx$ the
comoving Eulerian coordinates, from which physical distances may be
obtained according to the law $\bfr = a(t)\bfx$, $a(t)$ being the
expansion scale factor and $t$ the standard cosmic time.

According to the Lagrangian point of view, the path of each fluid element
is followed during its evolution. A proper observer may label each
neighbouring fluid particle by e.g. its initial (comoving) coordinate, say
$\bfq \equiv \bfx_{\circ}$. At time $t$, the (Eulerian) position of the
$\bfq$--particle will be
\be
\bfx = \bfx(\bfq,t) \equiv \bfx_L\;.
\ee
Here the only independent variables are the labels $\bfq$ (apart from the
time $t$) which therefore play the role of $spatial$ coordinates in
the Lagrangian $\{\bfq\}$--space. The vector $\bfx$, which in the
Eulerian picture is an independent variable, is now introduced
as a new real dynamical field: in the Lagrangian $\{\bfq\}$--space,
the trajectories of the mass elements are fully described by the dynamical
maps $\bfx(\bfq,t)$, starting from the initial positions
$\bfq$. The definition (1) implicitly
assumes that there is a one--to--one correspondence between the
Eulerian coordinate $\bfx$ and the Lagrangian coordinate $\bfq$: this
is certainly the case for a cold non collisional fluid, at least until
the stage of caustic formation (see e.g. Shandarin \& Zel'dovich 1989).
Mathematically, this is equivalent to the statement
that the determinant $J$ of the Jacobian of the
map $\bfq \rightarrow \bfx(\bfq,t)$ is non--singular,
\be
J(\bfq, t) \equiv {\rm det}\left(\f{\p\bfx }{\p \bfq}  \right) \neq 0\;,
\ee
the map $\bfx(\bfq,t)$ being thus reversible to $\bfq(\bfx,t)$.
Obviously, during the highly non--linear evolution, many particles
coming from very different original positions will tend to arrive
at the same Eulerian place: infinite--density regions (caustics) will form
in Eulerian space and the map from Lagrangian to Eulerian space
becomes singular (Shandarin \& Zel'dovich 1989; see also the discussion
in Kofman \etal 1994).
This caustic formation process enormously limits the predictive
power of any (perturbative) Eulerian method, in that the main
requirement for it to work is the very restrictive condition
$|\de|\ll 1$. This is possibly the strongest reason for preferring
the alternative Lagrangian picture: in this description
indeed the density is not a dynamical variable and is fully integrated.

Following Zel'dovich (1970a, b), all Eulerian fields can be
represented in terms of the only dynamical field $\bfx_L$ -- or its
derivatives. Thus, the peculiar velocity $\bfu$, the peculiar acceleration
$\bff$ and the density $\rho$ are respectively given by
\be
\f {d\bfx(\bfq,t)}{dt} = \bfu(\bfq,t)\,a(t)^{-1}\;,
\ee
\be
\f{d^2\bfx(\bfq,t)}{dt^2} = [\bff(\bfq,t) - 2H\bfu(\bfq,t)]\,a(t)^{-1}\;,
\ee
\be
1+ \de[\bfx(\bfq,t), t] = J(\bfq,t)^{-1}\;,
\ee
the quantity $H$ being the Hubble constant and
$\de = (\rho - \rho_b)/\rho_b$ the density
fluctuation about the density background value $\rho_b(t)$. The operator
$d/dt$ is the usual convective Lagrangian time derivative which
follows the mass element,
$\f{d}{dt} \equiv \f{\p}{\p t}|_{\bfq} = \f{\p}{\p t}|_{\bfx} +
\f{d\bfx}{dt}\cdot\nabla_{\bfx}\,$. Recall that the operator $\f{d}{dt}$
does not commute with the Eulerian nabla operator $\nabla_{\bfx}$.

Equation (4) corresponds to the (comoving) Euler equation and eq.(5) to the
continuity equation for the system. Furthermore, the
equations (3) and (4) -- governing the evolution
of the field $\bfx_L = \bfx(\bfq, t)$ -- may be thought as definitions
of the velocity $\bfu$ and the acceleration $\bff$. The particular
form of the mass conservation (4) derives from the fact that the observer
follows the mass elements during their motion and that the initial
matter distribution is assumed homogeneous.

The gravitational interaction among the mass elements is then introduced,
requiring that the peculiar acceleration field $\bff$ is induced
by the density fluctuations $\de$ through
the Poisson equation, $\nabla \cdot \bff = - 4\pi G a \rho_b\,\de$.
For our purposes, however, the best form of the
Poisson equation is the one obtained introducing the time variable
(Doroshkevich, Ryabenki \& Shandarin 1973; Shandarin 1980)
\be
\tau \equiv \sqrt{-k}\,(1-\Om)^{-1/2}\;,
\ee
where $k=-1$ for open universes and $k = 1$ for closed universes.
The case $\Om = 1 ~(k=0)$ is a singular point for the transformation
(6), and, in that case we take $\tau \equiv t^{-1/3}$.
As we shall see below, the main advantage in using $\tau$
instead of the standard cosmic time $t$ is that the dynamical
equations for the higher--order Lagrangian modes are considerably
simplified, since the rescaled peculiar acceleration $\bfg$ is now
simply given by
\be
\f{d^2\bfx}{d\tau^2} = \bfg\;.
\ee
Finally, gravity is
introduced through the equation
\be
\nabla_{\bfx}\cdot\bfg = -\al(\tau)\,\delta\;,
\ee
where the
quantity $\al(\tau)\equiv 6/(\tau^2+k)$ (Shandarin 1980).
The time variable $\tau$ has been recently used also in BJCP and
Lachi\`eze--Rey
(1993a,b).

Alternatively, if the original $\bff$ is expressed in terms of the
gravitational potential $\Phi$, i.e. $\bff \equiv - a^{-1}\nabla\Phi$, then,
because of (8),
\be
\nabla_{\bfx}^2\phi = \al(\tau)\,\de\;,
\ee
once the rescaled potential
$\phi\equiv(2a/\Om_{\circ} a_{\circ}\dot{a}_{\circ}^2)^2\,\Phi$ such
that $\bfg = -\nabla\phi$ has been introduced.

Finally, if one assumes irrotational motions in Eulerian space, which is
a plausible hypothesis for a collisionless (cold) dust, then
the ``irrotationality" condition may be expressed according to the
relation
\be
\nabla_{\bfx}\wedge\bfu = {\bf 0}\;.
\ee

Summarizing: the Lagrangian system of equations governing
a gravitating collisionless fluid is given by eqs.(3) and (4) -- the
Euler equation -- for the field $\bfx(\bfq,t)$, the mass conservation
relation (5) and the Poisson equation.
\section{Lagrangian Equations for the Displacement}
To describe the departure of the mass elements from the initial
positions $\bfq$ one usually introduces the {\it displacement vector}
$\bfS$ such that
\be
\bfx(\bfq,\tau) \equiv \bfq + \bfS(\bfq,\tau)\;.
\ee
It is clear from this definition that the motion of the fluid
elements may be completely described in terms of the
displacement $\bfS$, since the latter fully characterizes the map
(1) between the Eulerian and the Lagrangian coordinates.
In terms of $\bfS$, the Euler equation and the continuity equation
may be written, respectively, as
\be
\f{d^2\bfS(\bfq,\tau)}{d\tau^2} = \bfg[\bfx(\bfq,\tau),\tau]\;,
\ee
\be
1+\de[\bfx(\bfq,\tau),\tau] = {\rm det} (I + \calS)^{-1}\;.
\ee
Here $I = {\rm diag}(1,1,1)$ and $\calS$ is a $3\times 3$ matrix
whose elements are $S_{\al\beta}\equiv \p S_{\al}/\p q_{\beta}$, also
called the {\it deformation tensor}. In general the deformation tensor is
not symmetric, i.e. $S_{\al\beta}\neq S_{\beta\al}$: $S_{\al\beta}$ is
symmetric iff the
displacement $\bfS$ is potential in the Lagrangian space.
The Eulerian ``irrotationality" condition (10) --
if taken into account -- and the Poisson equation (8) may be written as
\be
\nabla_{\bfx}\wedge\dot{\bfS} = {\bf 0}\;,
\ee
\be
\nabla_{\bfx}\cdot\ddot{\bfS} = -\al(\tau)\,\de[\bfx(\bfq,\tau),\tau]\;,
\ee
the velocity field being defined by the relation
$d\bfx/d\tau=a(\tau)\,\bfu(\bfx, \tau)$.
The second equation clearly shows how the trajectories $\bfS$ are
deformed during the time evolution by the density perturbation
$\de[\bfx(\bfq,\tau),\tau]$. The dot indicates the
operator $d/d\tau$. However, we cannot say that
the previous equations are completely written in the Lagrangian
$\{\bfq\}$--space: indeed the operator $\nabla_{\bfx}$ does not
act on the Lagrangian coordinates $\bfq$. The exact way in which
the differentiation with respect to the Eulerian position $\bfx$
is translated -- through the map (11) -- into differentiation
with respect to the Lagrangian position $\bfq$ is displayed in
the following relation:
\be
J \,\f{\p}{\p x_{\beta}} =
\left[ (1+\nabla\cdot\bfS)\,\de_{\al\beta} - S_{\al\beta}
+ S^C_{\al\beta} \right] \f{\p}{\p q_{\al}}\;,
\ee
where now $\nabla \equiv \nabla_{\bfq}$. The symbol $\de_{\al\beta}$
indicates the Kronecker tensor and the quantity $S^C_{\al\beta}$
is an element of the cofactor matrix $\calS^C$. Summation over repeated
Greek indices (where $\al = 1, 2, 3$) is understood. We derive the
above relation in detail in Appendix A.

The Newtonian Lagrangian equations for the collisionless fluid
finally become
\be
\eps_{\al\beta\gamma}\,
\left[ (1+\nabla\cdot\bfS)\,\de_{\beta\sigma} - S_{\beta\sigma}
+ S^C_{\beta\sigma} \right] \dot{S}_{\gamma\sigma} = 0\;,
\ee
and
\be
\left[ (1+\nabla\cdot\bfS)\,\de_{\al\beta} - S_{\al\beta}
+ S^C_{\al\beta} \right] \ddot{S}_{\beta\al}
=  \al(\tau)[J(\bfq,\tau)-1]\;,
\ee
where $\eps_{\al\beta\gamma}$ is the totally antisymmetric Levi--Civita
tensor of rank three, $\eps_{123}\equiv 1$.

We call the latter equation the {\it Lagrangian Poisson equation}.
The irrotationality condition in Lagrangian space (17) and the Lagrangian
Poisson equation (18) dynamically constrain (up to a constant vector)
the field $\bfS$. They are
the (closed set of) $general$ dynamical equations for the
displacement vector $\bfS$
describing the motion of a collisionless fluid in the
Lagrangian $\{\bfq\}$--space, embedded in an arbitrary
non--flat universe and subject to the Newtonian gravitational influence of
the mass fluctuations $J^{-1}-1$. We stress that eqs.(17) and (18)
are non--linear and non--local in the displacement $\bfS$ (see the
discussion in Kofman \& Pogosyan 1994). Analogous equations, although
using a very different tensorial notation, are analyzed in Buchert (1989).

Some remarks are appropriate. Eq.(17) does not mean that the motion
is potential in the Lagrangian space, since this would correspond to the
condition $\eps_{\al\beta\gamma}\,\dot{S}_{\beta\gamma} = 0$.
On the contrary, we can surely
state from (17) that the motion is vortical in Lagrangian space and that,
however, the departure from the irrotationality of the Lagrangian
peculiar velocity $\bfu$ is gravitationally induced only at higher
order in the displacement $\bfS$, as may be clearly seen if one writes
$\eps_{\al\beta\gamma}\,\dot{S}_{\beta\gamma} =
- \eps_{\al\beta\gamma}\,
\dot{S}_{\beta\sigma}[(\nabla\cdot\bfS)\,\de_{\gamma\sigma}
- S_{\gamma\sigma} + S^C_{\gamma\sigma}]$. This implies that, at least
to the first order in $\bfS$, the condition for irrotationality in
Eulerian space means that the Lagrangian motion is also potential, as first
noted by Zel'dovich (1970a, b). The irrotationality problem in Lagrangian
coordinates has been recently addressed by Buchert (1992, 1994). We will
discuss again it below for a general nonflat model. Considering now the
Lagrangian Poisson equation, we stress the fact that only the left hand side
of eq.(18) is the `dynamical' part, containing the time derivatives
of the field $\bfS$: note that a term proportional to $\dot{\bfS}$ is
absent; furthermore, the Lagrangian Poisson equation
is $intrinsically$ non--linear in the density field, unlike the Eulerian
Poisson equation, as it may be seen from the relation
$J-1 = \f{1}{\de + 1} - 1$, where $\de$ is fully integrated.

It is remarkable the fact that the equations (17) and (18)
hold for a generic non--flat model, in that $\Om$ -- although fundamental --
is a very poorly known cosmological parameter and, apart from theoretical
expectations (essentially due to the implications of inflation),
there is no definitive observational evidence that our
Universe is flat (Peebles 1991; Coles \& Ellis 1994).
Furthermore, eqs.(17) and (18) are manifestly
comoving, which is also convenient, because the overall expansion
is usually subtracted in analytical or numerical studies of
departures from the mean homogeneous and isotropic universe.
In Appendix A, we give more compact expressions for eqs.(17) and
(18).

The irrotationality condition and the
Lagrangian Poisson equation are exact equations in the
Lagrangian description. It is undoubtedly very difficult
to solve them in a rigorous way. A possible alternative
is to seek for approximate solutions: the standard technique
is to expand the trajectory $\bfS$ in a perturbative series,
the leading term being the linear
displacement which corresponds indeed to the Zel'dovich
approximation (see Moutarde \etal 1991).
To approximate the Lagrangian Poisson equation
implies that the gravitational interaction among the particles of
the fluid is described only approximately.

\section{Lagrangian Perturbative Approximation: Higher--Order Solutions}
We now solve the dynamical equations for the displacements $\bfS$
according to the following Lagrangian perturbative prescription:
\be
\bfS(\bfq,\tau) =
D(\tau)\,\bfS^{(1)}(\bfq)
+E(\tau)\,\bfS^{(2)}(\bfq)+ F(\tau)\,\bfS^{(3)}(\bfq) + \cdots\;.
\ee
Here $\bfS^{(1)}(\bfq)$ corresponds to the first--order approximation,
$\bfS^{(2)}(\bfq)$ to the second--order approximation, and so on: the
dynamics of the evolution constrains in general both the temporal
dependence as described by the functions $D, E, F,...$, and the
spatial displacements $\bfS^{(n)}(\bfq)$.

Recently Gramann (1993), to calculate an analytical expression relating the
density to the velocity in a self--gravitating system, used a (second--order)
perturbative expansion similar to the previous one, but with the assumption
that $E \equiv D^2$: we will discuss below why this is a very restrictive
hypothesis.
Furthermore, we anticipate here -- and we will demonstrate in Section 4.3 --
that the third--order solution $\bfS^{(3)}$ actually corresponds to three
modes: we maintain here the expression as given in eq.(19) for the sake of
simplicity.

We emphasize that the first-- and the second--order perturbative
solutions (and, as it will be shown, each mode of the third--order
solution) are explicitly written in (19) as separable with respect to the
temporal and spatial coordinates. This is not an $assumption$, being just a
$property$ of the perturbative Lagrangian description. Indeed, this can be
demonstrated to be just a direct consequence of the dynamical
Lagrangian equations: while this turns out to be trivial for the
first--order solution, it is not so e.g. for the second--order
one. In Appendix B, it is shown that the separable solution
$E(\tau)\bfS^{(2)}(\bfq)$ is the most general second--order
Lagrangian perturbative solution.

Conversely, in Catelan et al. (1995) the {\it non--separability} of the
corresponding higher--order Eulerian perturbative solutions is thoroughly
analyzed: the Eulerian perturbative solutions factorize in space and time
only in the very special case of the flat universe, and not in a generic
Friedmann model; intriguingly, it is also shown that the non--separability
of the Eulerian solutions is fully consistent with the separability
of the Lagrangian solutions, at least explicitly up to the
second--order.

Because of their convenience in deriving the subsequent equations, we define
the following scalars:
\be
\mu_1(\bfA) \equiv \nabla\cdot\bfA = A_{\al\al}\;,
\ee
\be
\mu_2(\bfA,\bfB) \equiv
\f{1}{2}\,[A_{\al\al}B_{\beta\beta} - A_{\al\beta}B_{\beta\al}]\;,
\ee
\be
\mu_2(\bfA) \equiv \mu_2(\bfA,\bfA)\;,
\ee
\be
\mu_3(\bfA) \equiv {\rm det}(A_{\al\beta})\;,
\ee
where $\bfA$ and $\bfB$ are generic functions of the Lagrangian coordinate
$\bfq$ and spatial derivatives are with respect to the variable $\bfq$.
Note that the functions $\mu_1$, $\mu_2$ and $\mu_3$ are linear, quadratic
and cubic in their arguments, respectively.
Furthermore, the following expression for the Jacobian determinant $J$ holds:
\be
J(\bfq,\tau) = 1 + \mu_1(\bfS) + \mu_2(\bfS) + \mu_3(\bfS)\;.
\ee
This is an exact relation for the displacement and it is cubic in $\bfS$,
justifying investigation of the third--order solution. According to
eqs.(12) and (19), the Lagrangian acceleration may be perturbatively
expressed as
\be
\bfg[\bfx(\bfq,\tau),\tau] =
\ddot{D}(\tau)\,\bfS^{(1)}(\bfq)+
\ddot{E}(\tau)\,\bfS^{(2)}(\bfq)+
\ddot{F}(\tau)\,\bfS^{(3)}(\bfq) + \cdots\;,
\ee
explicitly up to the third--order term: this expression corresponds to the
Euler equation. We now solve the Lagrangian
Poisson equation order--by--order.
\subsection{First--Order Solution: Zel'dovich Approximation}
We can easily find the first--order approximation truncating eq.(18)
accordingly to the linear terms,
$\ddot{S}_{\al\al} = \al(\tau)\,\mu_1(\bfS)\,$, to obtain, in terms of the
displacement $\bfS^{(1)}$,
\be
\ddot{D}(\tau)\,S^{(1)}_{\al\al} = \al(\tau)\,D(\tau)\,\mu_1(\bfS^{(1)})\;.
\ee
Given the definition of $\mu_1$, we immediately get
\be
\ddot{D}(\tau) - \al(\tau)\,D(\tau) = 0\;.
\ee
The two linearly independent solutions coincide with the
growing and decreasing modes of the linear density field: hereafter
we will consider only the growing mode $D_+ \equiv D$, since
any perturbative approach is consistently applicable to the
(mildly) non--linear regime, when the decreasing modes have already
become negligible. We report here the linear growing solution
\be
D(\tau)=1+3\,(\tau^2-1)\left[1+
\tau\,{\rm ln}\sqrt{\f{\tau-1}{\tau+1}}\,\right]\;,
\ee
where we remind that, in the case $k=1$,
${\rm ln}\left(\f{1-\tau}{1+\tau}\right) = 2i\,{\rm arctang}(i\tau)$.
Eq.(27) thus corresponds to the Zel'dovich approximation. Note that the
linear regime does not constrain the vector $\bfS^{(1)}$ at all, whose
particular form has to be ascribed to the chosen initial conditions.
Furthermore, the displacement $\bfS^{(1)}$ is potential in Lagrangian
space, since the irrotationality condition (17) in the linear regime reduces to
\be
\eps_{\al\beta\gamma}\, S^{(1)}_{\gamma\beta} = 0\;,
\ee
as already known (Zel'dovich 1970a, b).
One can thus define a potential $\psi^{(1)}(\bfq)$ such that
$\bfS^{(1)}(\bfq)\equiv \nabla\psi^{(1)}(\bfq)$, with $\psi^{(1)}$
the velocity potential in the Zel'dovich approximation (see the
discussion in Kofman 1991). As a consequence, the linear deformation tensor
is symmetric. The particular growing solution (28) reduces to
$D = \tau^{-2}$ in the case of an Einstein--de Sitter model, since
$\al = 6\tau^{-2}$.
\subsection{Second--Order Solution: BJCP approximation}
Retaining only the quadratic terms in the Lagrangian Poisson
equation and introducing the ansatz $\bfS = D\bfS^{(1)}+E\bfS^{(2)}$, one gets
to second--order
$$
\ddot{E}\,\mu_1(\bfS^{(2)}) + D\ddot{D}\,\mu_1(\bfS^{(1)})^2 -
D\ddot{D}\,S^{(1)}_{\al\beta}S^{(1)}_{\beta\al} =
\al E\,\mu_1(\bfS^{(2)}) +\al D^2\,\mu_2(\bfS^{(1)})\;,
$$
from which, using the first--order results,
\be
\left[\ddot{E}(\tau) -\al(\tau)E(\tau)\right]\mu_1(\bfS^{(2)}) =
-\al(\tau)\,D(\tau)^2\,\mu_2(\bfS^{(1)})\;.
\ee
This is a separable differential equation leading to the system of
equations (cf. BJCP 1992)
\\
\be
\left\{
\begin{array}{l}
\ddot{E}(\tau) -\al(\tau)E(\tau) =-\al(\tau)\,D(\tau)^2\;, \\ \\
\mu_1(\bfS^{(2)}) = \mu_2(\bfS^{(1)})\;.
\end{array}
\right.
\ee
\\
We stress that, unlike the first--order case, the second--order
approximation constrains both the time and the spatial
dependence of the solution.
The solution $E(\tau)$ of the temporal equation
has been found by BJCP (1992). Its generality is discussed in
Appendix B. We report here its explicit
expression:
\be
E(\tau) = - \f{1}{2} -
\f{9}{2}\,(\tau^2-1)\left\{1+\tau\,{\rm ln}\sqrt{\f{\tau-1}{\tau+1}}
+\f{1}{2}\left[\tau +(\tau^2-1)\,{\rm ln}\sqrt{\f{\tau-1}{\tau+1}}\,\right]^2
\right\}\;.
\ee
In BJCP, this solution has been applied to describe the
dependence of the skewness of the unfiltered density field on the
density parameter $\Om$, in that, near $\Om = 1$,
$E \approx - \f{3}{7}\Om^{-2/63}D^2$. For us it will be useful, because
we will be able to express the third--order solutions in terms of
the lower--order results.

The solution  of the second--order spatial equation in (31) may be
written as
\be
\bfS^{(2)} =
\f{1}{2}\,\left[\bfS^{(1)}\left(\nabla\cdot\bfS^{(1)}\right) -
\left(\bfS^{(1)}\cdot\nabla\right)\bfS^{(1)}\right] + \bfR^{(2)}\;,
\ee
where $\bfR^{(2)}(\bfq)$ is any divergence--free vector such that
$\nabla\wedge \bfS^{(2)} ={\bf 0}$: $i\!f$ the first--order Lagrangian
motion is assumed potential, this is indeed the result
one obtains from the irrotationality equation (17) once only the
second--order terms are retained
\be
\dot{E}(\tau)\left(\nabla\wedge\bfS^{(2)}\right)_{\al} =
- \dot{D}(\tau)D(\tau)\,\eps_{\al\beta\gamma}\,
S^{(1)}_{\beta\sigma}S^{(1)}_{\gamma\sigma} = 0\;,
\ee
where the last equality follows from the fact that the tensor
$S^{(1)}_{\beta\sigma}S^{(1)}_{\gamma\sigma}$
is symmetric. Thus, we have to conclude  that the gravitational
evolution does not induce vorticity in the second--order Lagrangian motion
at all, whatever the initial conditions are: note that
$\nabla\wedge\bfS^{(2)}={\bf 0}$ is a purely $spatial\,$ relation.
It appears clear from
(34) that the displacement $\bfS^{(2)}$ is potential once $\bfS^{(1)}$
is assumed to be potential too, and thus the second--order deformation
tensor is symmetric: a potential $\psi^{(2)}$ may be introduced, from
which one obtains $\bfS^{(2)}(\bfq) = \nabla\psi^{(2)}(\bfq)$. An
useful expression of the Fourier transform of $\psi^{(2)}$,
$\widetilde{\psi}^{(2)}$, in terms
of the first--order potential $\widetilde{\psi}^{(1)}$ may be found
in Appendix C: it will clearly result that an explicit expression
of the gauge--dependent vector $\bfR^{(2)}$ is, in practice, unnecessary:
we nevertheless emphasize that the  expression in (33) is the only
compatible with the irrotationality condition (34).
A restricted class of second--order irrotational solutions, for
which additional constraints on the initial conditions have to be fulfilled,
is discussed in  Buchert \& Ehlers (1993) and Buchert (1994); the same
constraints also allow the construction of local forms, as debated in
Buchert (1994), again in the context of the Einstein--de Sitter
cosmology. The solution in (32) reduces to the simple form
$E = -\f{3}{7}\tau^{-4}\propto D^2$ in the case of the flat model.
\subsection{Third--Order Solution}
Inserting the ansatz (19) in the Lagrangian Poisson equation,
after some algebra one obtains the third--order expression
$$
\ddot{F}\,\mu_1(\bfS^{(3)}) +
2(D\ddot{E}+\ddot{D}E)\,\mu_2(\bfS^{(1)},\bfS^{(2)}) +
3\ddot{D}D^2\,\mu_3(\bfS^{(1)}) \;\;\;\;\;\;\;\;\;\;\;\;\;\;\;\;\;
$$
$$
\;\;\;\;\;\;\;\;\;\;\;\;\;
\;\;\;\;\;\;\;\;\;\;\;\;\;= \al(\tau)\left[\,F\,\mu_1(\bfS^{(3)}) +
2DE\,\mu_2(\bfS^{(1)},\bfS^{(2)}) + D^3\,\mu_3(\bfS^{(1)}) \,\right]\;,
$$
where we used the relation
\be
S^C_{\al\beta}S^{(1)}_{\beta\al} =
D^2\,S^{(1)C}_{\al\beta}S^{(1)}_{\beta\al} = 3 D^2 \mu_3(\bfS^{(1)})\;.
\ee
On the light of the lower--order results, the former expression
becomes
$$
\left[\ddot{F}(\tau)-\al(\tau)F(\tau)\right]\mu_1(\bfS^{(3)})\;\;\;\;\;\;
\;\;\;\;\;\;\;\;\;\;\;\;\;\;\;\;\;\;\;\;\;\;\;\;
$$
\be
= -2\al(\tau)D(\tau)\left[E(\tau)-D(\tau)^2\right]
\mu_2(\bfS^{(1)},\bfS^{(2)})-2\al(\tau)D(\tau)^3\,\mu_3(\bfS^{(1)})\;.
\ee
Note that the mixed invariant $\mu_2$ couples the first-- and
second--order solutions $\bfS^{(1)}$ and $\bfS^{(2)}$, while the last
term $\mu_3$ is cubic in the argument $\bfS^{(1)}$. This fact
forces us, as suggested by Buchert (1994), to split the third--order
displacement $\bfS^{(3)}$ into two parts, one resulting from the interaction
among the linear perturbations, the second from the interaction between
the first-- and second--order perturbations:
\be
\bfS^{(3)}(\bfq) = \bfS^{(3)}_a(\bfq) + \bfS^{(3)}_b(\bfq)\;.
\ee
According to this ansatz, the dynamical equation (36) splits into the
systems
\\
\be
\left\{
\begin{array}{l}
\ddot{F}_a(\tau) -\al(\tau)F_a(\tau) =-2\al(\tau)\,D(\tau)^3\;, \\ \\
\mu_1(\bfS^{(3)}_a) = \mu_3(\bfS^{(1)})\;,
\end{array}
\right.
\ee
\\
and
\\
\be
\left\{
\begin{array}{l}
\ddot{F}_b(\tau) -\al(\tau)F_b(\tau) =-2\al(\tau)\,D(\tau)
\left[E(\tau)-D(\tau)^2 \right]\;, \\ \\
\mu_1(\bfS^{(3)}_b) = \mu_2(\bfS^{(1)},\bfS^{(2)})\;.
\end{array}
\right.
\ee
\\
Note that, since now $\bfS^{(2)}$ may be considered a potential
field, $\mu_2(\bfS^{(1)},\bfS^{(2)}) = \mu_2(\bfS^{(2)},\bfS^{(1)})$.

It clearly appears that the assumption $E\equiv D^2$ [see Gramann (1993)]
implies a neglect of the interaction mode between the first-- and
second--order perturbations, in that $E=D^2 \Longrightarrow F_b \equiv 0$.
This is in general a very restrictive hypothesis and possibly only
(high--resolution) $N$--body simulations might quantify the real loss
of accuracy at the third--order level.

Furthermore, Lachi\`eze--Rey (1993b) argues that one formal third--order
solution exists corresponding to the assumption that the deformation
tensor $\calS(\bfq,\tau)$ remains proportional, at each Lagrangian position,
to its initial value, being only multiplied by a scalar Lagrangian growth
factor. This is surely true, but the problem is now that such a
hypothesis notably restricts the form of the $spatial$ part of the
third--order solution: it is not difficult indeed to show that the
Lachi\`eze--Rey solution corresponds {\it to assuming} that, in our notation,
$\mu_3(\bfS^{(1)}) = \mu_2(\bfS^{(1)}, \bfS^{(2)})$
and, after that, $\mu_1(\bfS^{(3)}) = \mu_3(\bfS^{(1)})$. Of course
these conditions depend also on the particular chosen initial
configurations, but it is not clear to which type of realistic
physical situation they may be applied.

After these comments, we give now the third--order solutions by
quadrature:
\be
F_a(\tau) = -2\,D(\tau)\int^{\tau}d\tau_1\, D(\tau_1)^{-2}
\int^{\tau_1} d\tau_2\, \al(\tau_2)\,D(\tau_2)^4\;,
\ee
and
\be
F_b(\tau) = -2\,D(\tau)\int^{\tau}d\tau_1\, D(\tau_1)^{-2}
\int^{\tau_1} d\tau_2 \,\al(\tau_2)\,D(\tau_2)^2
\left[E(\tau_2) - D(\tau_2)^2 \right]\;.
\ee
In this way the two solutions $F_a(\tau)$ and $F_b(\tau)$ are given
in terms of the lower--order solutions $D(\tau)$ and $E(\tau)$:
explicit versions of (40) and (41) are  cumbersome, while
approximate expressions may be recovered in the limit e.g.
$\Om \rightarrow 1$ (Catelan 1995). In the case of an Einstein--de Sitter
model, we
find $F_a = -\f{1}{3}\tau^{-6}$ and $F_b = \f{10}{21}\tau^{-6} \propto D^3$:
the results in Buchert (1994) are particular cases of the general
solutions (40) and (41).

A solution $\bfS^{(3)}_a$ of (38) may be obtained recalling that, from
(35), $S^{(1)C}_{\al\beta}S^{(1)}_{\beta\al} = 3 \mu_3(\bfS^{(1)})$; thus,
since $\p_{\al}S^{(1)C}_{\al\beta}=0$, one gets
$\p_{\al}\left(S^{(3)}_{a\,\al}-
\f{1}{3}S^{(1)C}_{\al\beta}S^{(1)}_{\beta}\right) = 0$ and finally,
for each component,
\be
S^{(3)}_{a\,\al} = \f{1}{3}\,S^{(1)C}_{\al\beta}S^{(1)}_{\beta} +
R^{(3)}_{a\,\al}\;.
\ee
Here $\bfR^{(3)}_a(\bfq)$ is again a divergence--free vector such that
$\nabla\wedge\bfS^{(3)}_a={\bf 0}$. To understand why $\bfS^{(3)}_a$
is a potential vector it is indeed enough to isolate, in the
Lagrangian irrotationality condition (17), all the possible vortical
terms induced by the interactions among the first--order displacements:
it is not difficult to see that the only term
\be
\eps_{\al\beta\gamma}\,S^{(1)}_{\beta\sigma}
S^{(1)C}_{\gamma\sigma}=0\;,
\ee
where the equality is justified by the assumption that
the linear displacement is potential. Another way to describe the
irrotationality of $\bfS^{(3)}_a$ is to note the that third--order interactions
like $S^{(1)}_{\ast\ast}S^{(1)}_{\ast\ast}S^{(1)}_{\ast\ast}$ can be only
symmetric in $S^{(1)}_{\ast\ast}$ and not antisymmetric (here $\ast$ indicate
a generic Greek index), as any possible vortical--like interaction would be.

In a similar fashion, one can derive a solution $\bfS^{(3)}_b$ from (39),
assuming in particular that $\bfS^{(3)}_b$ is induced by symmetric
couplings between the first-- and the second--order solutions.
It can be written as
\be
\bfS^{(3)}_b = \f{1}{4}
\left[\bfS^{(1)}(\nabla\cdot\bfS^{(2)})-(\bfS^{(1)}\cdot\nabla)\bfS^{(2)}+
\bfS^{(2)}(\nabla\cdot\bfS^{(1)})-(\bfS^{(2)}\cdot\nabla)\bfS^{(1)}\right]
+\bfR^{(3)}_b\;,
\ee
where again $\bfR^{(3)}_b(\bfq)$
is a divergence--free vector such that $\nabla\wedge\bfS^{(3)}_b={\bf 0}$:
in fact, symmetric interactions between the first-- and second--order
modes cannot be vortical.

The solutions $\bfS^{(3)}_a$ and $\bfS^{(3)}_b$, being longitudinal, may be
written in terms of the respective potentials $\psi^{(3)}_a$ and
$\psi^{(3)}_b$,
namely $\bfS^{(3)}_a(\bfq)\equiv\nabla\psi^{(3)}_a(\bfq)$ and
$\bfS^{(3)}_b(\bfq)\equiv\nabla\psi^{(3)}_b(\bfq)$. In Appendix C, the
expressions of the Fourier components $\widetilde{\psi}^{(3)}_a$ and
$\widetilde{\psi}^{(3)}_b$ as functions of the Zel'dovich potential
$\widetilde{\psi}^{(1)}$ are given. The significance of the vectors
$\bfR^{(3)}_a$ and $\bfR^{(3)}_b$ is perfectly equivalent to that
of the vector $\bfR^{(2)}$ for the second--order solution $\bfS^{(2)}$.

One now has to make the following inquiry: does
any $antisymmetric\,$ coupling between the
Zel'dovich and the BJCP solutions exist at the
third--order level? If it exists, then a third--order
transverse component $\bfT$ arises in the Lagrangian motion of
the fluid mass elements, and this is the $only$ vortical component
which can be induced in the framework of the third--order approximation.
Note that such a type of component cannot modify the amplitude
of the density fluctuation field.

Let us suppose therefore that such a transverse mode actually exists: if
the original ansatz (19) is now improved to
\be
\bfS(\bfq,\tau) = \bfS_{\parallel} + \bfS_{\perp} =
\bfS_{\parallel} + F_c(\tau)\,\bfT(\bfq)\;,
\ee
where we recall that
$\bfS_{\parallel}\equiv
D\bfS^{(1)}+E\bfS^{(2)}+F_a\bfS^{(3)}_a+F_b\bfS^{(3)}_b$,
it follows that, in terms of $\bfT$, the Lagrangian irrotationality
condition becomes
\be
\dot{F}_c(\tau)
\left(\nabla\wedge\bfT\right)_{\al} =
\left[D(\tau)\dot{E}(\tau) -\dot{D}(\tau)E(\tau) \right]\,
\eps_{\al\beta\gamma}\,S^{(1)}_{\beta\sigma}S^{(2)}_{\sigma\gamma}\;.
\ee
We derive eq.(46) explicitly in Appendix D.
Again, this equation may be split into a temporal and a spatial part:
\\
\be
\left\{
\begin{array}{l}
\ddot{F}_c(\tau) = -\al(\tau)\,D(\tau)^3 \;, \\ \\
\left(\nabla\wedge\bfT\right)_{\al} =
\eps_{\al\beta\gamma}\,S^{(1)}_{\beta\sigma}S^{(2)}_{\sigma\gamma} \;.
\end{array}
\right.
\ee
\\
The growth factor $F_c(\tau)$ may be explicitly written by quadrature
\be
F_c(\tau) = -\int^{\tau}d\tau_1\int^{\tau_1}d\tau_2\,\al(\tau_2)
\,D(\tau_2)^3\;.
\ee
This solution reduces to the simple $F_c = - \f{1}{7}\tau^{-6}$ in
the case of an Einstein--de Sitter universe.

{}From the second equation in (47), we note -- consistently -- that the
antisymmetric part of the third--order interaction between
the modes $\bfS^{(1)}$ and $\bfS^{(2)}$ generates
the transverse component
\be
\bfT(\bfq) = \f{1}{2}\left[(\bfS^{(1)}\cdot\nabla)\bfS^{(2)}
-(\bfS^{(2)}\cdot\nabla)\bfS^{(1)} \right] + \nabla \varphi(\bfq)\;.
\ee
In Appendix C the Fourier components $\widetilde{T}_{\al}$ are
explicitly given.
The term $\nabla\varphi$ is such that $\nabla\cdot\bfT = 0$ or,
equivalently, the transverse component $\bfT$ can $now$ be described
as the curl of a vector potential, $\bfT \equiv \nabla \wedge {\bf A}$,
which cannot in general be removed by a suitable gauge--fixing of the
initial conditions: one can say that the Lagrangian motion is no
longer purely potential from third--order onward.

This fact was first discovered by Buchert (1994) within
the Einstein--de Sitter model, and therefore seems
a universal feature of the Lagrangian motion, independently of the
value of the density parameter $\Om$ (i.e. of the underlying
model of universe). However, it is important to stress again that
the displacements $\bfS^{(3)}_a$ and $\bfS^{(3)}_b$ are in general purely
potential just because they originate through symmetric couplings
among the lower--order longitudinal perturbations; also, the
transverse displacement $\bfT$ is purely vortical because it originates
through the antisymmetric coupling between the first-- and
second--order solutions, $independently\,$ of the
peculiar initial conditions one is picking out. We summarize
the results of this work in the next section.
\section{Summary and Conclusions}
In this work we studied, in the Lagrangian description, the behaviour of a
self--gravitating collisionless fluid embedded in a generic non--flat
expanding universe. We assumed the Newtonian
limit and zero cosmological constant. We have mainly focussed on
the formal aspects of this kind of analysis, also proposing a new
Lagrangian formalism. In particular, we derived the general equations,
in the forms given in (17) and (18), governing the Lagrangian motion of
the mass elements in a universe with arbitrary density parameter
$\Om$ and directly in comoving coordinates. We consider this kind of
derivation one of the main results of this work. An alternative derivation
and notation may be found in
Buchert (1989). Since it is in practice very difficult to
find exact solutions of the foregoing general equations, we solved
them according to a Lagrangian perturbative approach, namely we
sought approximate solutions $\bfS = \sum \bfS^{(n)}$ about
the linear displacement $\bfS^{(1)}$ as pioneered by Moutarde \etal (1991).
Our formalism
enables one to easily recover the already known lower--order solutions,
in particular the linear Zel'dovich approximation (Zel'dovich 1970a, b)
and the second--order BJCP approximation (Bouchet \etal 1992). Then, we
explicitly worked out the third--order solution, generalizing to
an arbitrary Friedmann universe the recent results obtained
by Buchert (1994) in the context of the Einstein--de Sitter cosmology.
The question of the irrotationality in the Lagrangian space has
been analyzed too. In particular, we found that the spatial solutions
$\bfS^{(2)}$, $\bfS^{(3)}_a$, $\bfS^{(3)}_b$ are purely potential
for any acceptable initial conditions -- the only underlying hypothesis
being that the linear displacement is longitudinal -- as
one finds if one carefully applies, order--by--order, the irrotationality
condition (17). Similarly, the third--order transverse component $\bfT$ is
purely vortical for any realistic initial conditions. Obviously, the reason
of the existence of the transverse component $\bfT$ is that the
transformation from Eulerian to Lagrangian coordinates
$\bfx \rightarrow \bfq$ is in general a non--Galilean transformation,
and fictitious forces are induced in the Lagrangian description: no
new physics can indeed appear.

To avoid an exceeding proliferation of formulae in the main text of the
article, we performed in Appendix C the Fourier analysis of the
higher--order Lagrangian solutions: such a results can be important for
numerical and practical applications. Specifically, we found that the
third--order
dynamics is fully described in terms of the tetra--potential
$\Psi\equiv (\psi^{(3)}_a, \psi^{(3)}_b, A_1, A_2)$, since only two of the
three components of the vector potential $\bfA$ are physically significative.

Our results offer a tool with which to follow the dynamics of the formation
of the structures in the universe, as resulting
from the non--linear gravitational instability. \\

\noindent {\bf Acknowledgments}~ I am indebted to Sabino Matarrese for
many useful suggestions and to Cedric Lacey for enlightening remarks.
James Binney, Peter Coles, George Efstathiou, Francesco Lucchin,
Lauro Moscardini, and Mikel Susperregi are also
warmly thanked for discussions. I am very grateful to Sergei Shandarin for
his precious comments and encouragements and to Thomas Buchert for clarifying
in detail his argumentation about the construction of higher--order
local Lagrangian solutions. Lauro Moscardini is also thanked for technical
help, and Francis Bernardeau for his very constructive criticisms.
A meticulous anonymous referee improved the presentation of this article.
This work was supported by EEC Human Capital and Mobility Programme and by
Fondazione Angelo Della Riccia.\\

\noindent To Elena, she had the time of knowing it.

\vspace{1cm}

\noindent {\bf Note added in proof}~ After the submission
of this article, Thomas Buchert addressed our attention on the
Appendix A in Buchert (1989) where, employing the same
temporal parameter in (6), an alternative expression of the
equation (18) is given. The temporal coordinate (6) has been
recently used also in Bharadwaj, S. 1994, ApJ, 428, 419 and
Bouchet, F.R., \etal 1994, A\&A, submitted.

\section*{References}
\begin{trivlist}

\item[] Bagla, J.S., \& Padmanabhan, T. 1994, MNRAS, 266, 227

\item[] Bernardeau, F. 1992, ApJ, 392, 1

\item[] Bernardeau, F. 1994, ApJ, 433, 1

\item[] Bernardeau, F., Singh, T.P., Banerjee, B., \& Chitre, S.M. 1994,
        MNRAS, 269, 947

\item[] Brainerd, T.G., Scherrer, R.J., \& Villumsen, J.V. 1993, ApJ, 418, 570

\item[] Bouchet, F.R., Juszkiewicz, R., Colombi, S., \& Pellat, R. 1992, ApJ,
	394, L5 (BJCP)

\item[] Buchert, T. 1989, A\&A, 223, 9

\item[] Buchert, T. 1992, MNRAS, 254, 729

\item[] Buchert, T. 1994, MNRAS, 267, 811

\item[] Buchert, T. \& Ehlers, J. 1993, MNRAS, 264, 375

\item[] Catelan, P. 1995, in preparation

\item[] Catelan, P., Lucchin, F., Matarrese, S., \& Moscardini, L. 1995,
        MNRAS, in press

\item[] Catelan, P., \& Moscardini, L. 1994a, ApJ, 426, 14

\item[] Catelan, P., \& Moscardini, L. 1994b, ApJ, 436, 5

\item[] Coles, P., \& Ellis, G.F.R., 1994, Nature, 370, 609

\item[] Coles, P., Melott, A.L., \& Shandarin, S. 1993, MNRAS, 260, 765

\item[] Coles, P., Moscardini, L., Lucchin, F., Matarrese, S., \& Messina, A.
        1993, MNRAS, 264, 749

\item[] Doroshkevich, A.G., Ryabenki, V.S., \& Shandarin, S.F. 1973,
        Astrofizika, 9, 257 (1975, Astrophysics, 9, 144)

\item[] Fry, J.N. 1984, ApJ, 279, 499

\item[] Goroff, M.H., Grinstein, B., Rey, S.--J., \& Wise, M.B. 1986, ApJ,
	311, 6

\item[] Gramann, M. 1993, ApJ, 405, L47

\item[] Juszkiewicz, R., Bouchet, F.R., \& Colombi, S. 1993, ApJ, 412, L9

\item[] Kofman, L. 1991, in {\it Primordial Nucleosynthesis and Evolution of
        the Early Universe}, ed. Sato, K., Dordrecht: Kluwer Academic

\item[] Kofman, L., Bertschinger, E., Gelb, M.J., Nusser, A., \& Dekel, A.
        1994, ApJ, 420, 44

\item[] Kofman, L., \& Pogosyan, D. 1994, ApJ, submitted

\item[] Lachi\`eze--Rey, M. 1993a, ApJ, 407, 1

\item[] Lachi\`eze--Rey, M. 1993b, ApJ, 408, 403

\item[] Matarrese, S., Lucchin, F., Moscardini, L., \& Saez, D. 1992, MNRAS,
        259, 437

\item[] Matarrese, S., Pantano, O., \& Saez, D. 1994a, Phys. Rev. Lett., 72,
320

\item[] Matarrese, S., Pantano, O., \& Saez, D. 1994b, MNRAS, in press

\item[] Melott, A.L., Lucchin, F., Matarrese, M., \& Moscardini, L. 1994,
        ApJ, 422, 430

\item[] Moutarde, F., Alimi, J.--M., Bouchet, F.R., Pellat, R., \&
        Ramani, A. 1991, ApJ, 382, 377

\item[] Munshi, D., Sahni, V.,\& Starobinsky, A.A. 1994, ApJ, submitted

\item[] Munshi, D., \& Starobinsky, A.A. 1994, ApJ, 428, 433

\item[] Nusser, T.A., \& Dekel, A. 1992, ApJ, 391, 443

\item[] Peebles, P.J.E. 1980, The Large Scale Structure of the Universe
        (Princeton: Princeton Univ. Press)

\item[] Peebles, P.J.E. 1991, in {\it Observational Tests of Inflation},
        ed. Shanks, T., Dordrecht: Kluwer Academic

\item[] Shandarin, S.F. 1980, Astrofizika, 16, 769 (1981, Astrophysics, 16,
439)

\item[] Shandarin, S.F., \& Zel'dovich, Ya.B. 1989, Rev. Mod. Phys., 61, 185

\item[] Zel'dovich, Ya.B. 1970a, A\&A, 5, 84

\item[] Zel'dovich, Ya.B. 1970b, Astrophysics, 6, 164

\end{trivlist}

\newpage

\bc
\section*{Appendix A}
\ec

In the first appendix we explicitly show how to obtain
the general Lagrangian equations (17) and (18).
Let us start by deriving the Lagrangian Poisson equation: from
eq.(15), this can be written in the form\be
J(\bfq, \tau) \f{\p \ddot{S}_{\al}(\bfq,\tau)}{\p x_{\al}} =
\al(\tau)[J(\bfq,\tau) - 1]\;,
\ee
where $J$ is the determinant of the Jacobian
$\left(\f{\p\bfx}{\p\bfq}\right)$,
and $\bfx =\bfq + \bfS(\bfq,\tau)$.
To derive (18), all we need is to show that the following
relation holds
\be
J \f{\p}{\p x_{\al}} =
\left[ (1+S_{\gamma\gamma})\,\de_{\beta\al} - S_{\beta\al}
+ S^C_{\beta\al} \right] \f{\p}{\p q_{\beta}}\;.
\ee
This indeed corresponds to the implicit expression
\be
\f{\p}{\p x_{\al}} = \f{\p q_{\beta}}{\p x_{\al}} \,\f{\p}{\p q_{\beta}}
= \left(\f{\p x_{\al}}{\p q_{\beta}}\right)^{-1} \f{\p}{\p q_{\beta}}\;.
\ee
The quantities $\left(\f{\p x_{\al}}{\p q_{\beta}}\right)^{-1}$
are elements of the matrix
$\left(\f{\p\bfx}{\p\bfq}\right)^{-1}=\left(\f{\p\bfq}{\p\bfx}\right)$.
Since, explicitly,
\\
\be
\left(\f{\p\bfx}{\p\bfq}\right)=
\left(
\begin{array}{ccc}
1+S_{11} & S_{12} & S_{13} \\
S_{21} & 1+S_{22} & S_{23} \\
S_{31} & S_{32} & 1+ S_{33}
\end{array}
\right)\;,
\ee
\\
we obtain, in matrix notation,
\be
\left(\f{\p\bfx}{\p\bfq}\right)^{-1} =
\f{1}{J}\left\{(1+\nabla\cdot\bfS)I - \calS + \calS^C \right\}\;,
\ee
where $I={\rm diag}(1,1,1)$ is the identity, $\calS \equiv (S_{\al\beta})$
is the deformation matrix and $\calS^C$ is the cofactor matrix we
usefully write below:
\\
\be
\calS^C \;\;\;\equiv\;\;\;
\left(
\begin{array}{ccc}
S_{22}S_{33}-S_{32}S_{23}\;\;\;\;&\;\;\;\;
S_{32}S_{13}-S_{12}S_{33}\;\;\;\;&\;\;\;\;S_{12}S_{23}-S_{22}S_{13}\\ \\
S_{31}S_{23}-S_{21}S_{33}\;\;\;\;&\;\;\;\;
S_{11}S_{33}-S_{31}S_{13}\;\;\;\;&\;\;\;\;S_{21}S_{13}-S_{11}S_{23}\\ \\
S_{21}S_{32}-S_{31}S_{22}\;\;\;\;&\;\;\;\;
S_{31}S_{12}-S_{11}S_{32}\;\;\;\;&\;\;\;\;S_{11}S_{22}-S_{21}S_{12}
\end{array}
\right)\;.
\ee
\\
With these results one now immediately recovers the
Lagrangian Poisson equation from the original eq.(15).

In exactly the same way one can obtain the
irrotationality condition in the Lagrangian space from the
Eulerian eq.(10):
\be
\eps_{\al\beta\gamma}\,\f{\p u_{\gamma} }{\p x_{\beta}} = 0\;.
\ee
Recalling eqs.(11), (12) and (52), this becomes
\be
\eps_{\al\beta\gamma}\left(\f{\p x_{\beta}}{\p q_{\sigma}} \right)^{-1}
\f{\p\dot{S}_{\gamma}}{\p q_{\sigma}} = 0\;.
\ee
If now the relation (51) is inserted in the last equation, we get
the general equation (17).

To conclude we propose here even more compact forms for the
irrotationality condition (17) and the
Lagrangian Poisson equation (18): in fact, defining the
cofactor element of $x_{\al\beta}$
\be
x^C_{\al\beta} \equiv J\, x^{-1}_{\al\beta} =
\left[ (1+\nabla\cdot\bfS)\,\de_{\al\beta} - S_{\al\beta}
+ S^C_{\al\beta} \right]\;,
\ee
where
$x^{-1}_{\al\beta}\equiv\left(\f{\p x_{\al}}{\p q_{\beta}}\right)^{-1}$,
we may finally write
\be
\eps_{\al\beta\gamma}\,x_{\beta\sigma}^C\,\dot{x}_{\gamma\sigma} = 0\;,
\ee
\be
x_{\al\beta}^C\,\ddot{x}_{\beta\al} = \al(\tau)[J-1]\;,
\ee
respectively. We have nevertheless preferred to retain less compact
and elegant versions in the main text, because the dependence on the
displacement $\bfS$ and its derivatives is there explicitly shown.

\bc
\section*{Appendix B}
\ec
In this appendix it is demonstrated that the $separable$ second--order
solution $E(\tau)\bfS^{(2)}(\bfq)$ is the most general solution of the
Lagrangian fluid equations, once only the second--order terms are retained.
Let us suppose that, {\it ab absurdo},
the second--order solution is non--separable (and longitudinal for
simplicity), namely
$D(\tau)^2\,\nabla\Phi_2(\bfq, \tau)$: the factorization of the term
$D^2$ does not alter the demonstration. The first order
solution is given in Section 4.1. From the Lagrangian Poisson
equation (18), one gets to second--order
\be
D^2\p^2_{\tau}\fPhi_2 +4D\dot{D}\p_{\tau}\fPhi_2 +
\Big(2\dot{D}^2 + 2D\dot{D}-\al(\tau)D^2\Big)\fPhi_2 = -\al(\tau)D^2\fcalP_2\;,
\ee
where the tilde ``$\tilde{\;\;}$'' indicates the Fourier transformed quantities
(see next appendix). The function $\fcalP_2$, whose explicit expression is
superfluous to give here (the interested reader is addressed to the next
appendix), is defined by the
relation $\fcalP_2 \equiv -p^{-2}\widetilde{\mu_2(\bfS^{(1)})}$:
the important point is to note that $\calP_2$ depends $only$
on the spatial variable: $\fcalP_2=\fcalP_2(\bfp)$, where
$\bfp$ indicates the comoving Lagrangian wavevector, and
$p\equiv |\bfp|$. Now it is
easy to verify that, if the function $B(\tau)$ satisfying the
differential equation $K^2\ddot{B}+4K\dot{B}+(2+\al K^2)B=1$,
where $K\equiv D/\dot{D}$,
is introduced, the function $\fPhi_2$ may be recast in the
separable form
\be
\fPhi_2(\bfp, \tau) = [2B(\tau)-1]\,\fcalP_2(\bfp) \;.
\ee
Then the function $E$ so defined, $E(\tau)\equiv D(\tau)^2[2B(\tau)-1]$,
satisfies the differential equation $\ddot{E}-\al(\tau)E=-\al(\tau)D(\tau)^2$,
which is the first equation in (31).
This concludes the demonstration. Similar considerations may
be extended to higher--order modes (Buchert, private communication;
Ehlers \& Buchert, 1995, in preparation).

\bc
\section*{Appendix C}
\ec
In this third appendix, we perform the complete Fourier analysis of the
Lagrangian motion described in the main text. The final results are useful
for practical and numerical applications.

Let us indicate by $\bfp$ the comoving Lagrangian wavevector. The $n$th--order
displacement potential $\psi^{(n)}(\bfq)$ may be written as a Fourier integral,
\be
\psi^{(n)}(\bfq)=\f{1}{(2\pi)^3}\int \,d\bfp
\,\widetilde{\psi}^{(n)}(\bfp)\,{\rm e}^{i\,\bfp\cdot\bfq}\;,
\ee
where it is understood that we restrict, in this appendix, to the cases
$n = 1,2,3$. Observing that
$i p_{\al}\widetilde{\psi}^{(n)}(\bfp)=\widetilde{S}_{\al}^{(n)}(\bfp)$,
it is immediate to obtain from the solution (33) that
\be
\widetilde{\psi}^{(2)}(\bfp)
=-\f{1}{p^2}\int \f{d\bfp_1 d\bfp_2}{(2\pi)^6}
\Big[(2\pi)^3 \delta_D(\bfp_1+\bfp_2 - \bfp)\Big]\,
\kappa^{(2)}(\bfp_1,\bfp_2)
\,\widetilde{\psi}^{(1)}(\bfp_1)\,\widetilde{\psi}^{(1)}(\bfp_2)\;,
\ee
where we have defined the kernel
\be
\kappa^{(2)}(\bfp_1,\bfp_2)\equiv
\f{1}{2}\,\Big[\,p_1^2\,p_2^2 - (\bfp_1\cdot\bfp_2)^2\Big] =
\f{1}{2}\,\Big(\,p_1\,p_2\, {\rm sin}\theta_{12}\Big)^2\;,
\ee
being e.g. $p \equiv |\bfp|$ and
$\theta_{12} \equiv {\rm arcos}(\bfp_1\cdot\bfp_2/p_1\, p_2)$ the angle between
the vectors $\bfp_1$ and $\bfp_2$; the presence of the Dirac--function
$\delta_D$ comes from momentum conservation in Fourier space. An alternative
simpler expression of $\widetilde{\psi}^{(2)}(\bfp)$ may be obtained performing
one integration: the result may be written as follows,
\be
\widetilde{\psi}^{(2)}(\bfp)= -\f{1}{p^2}\int \f{d\bfp'}{(2\pi)^3}
\,\,\kappa^{(2)}(\bfp,\bfp')\,\,
\widetilde{\psi}^{(1)}(\bfp')\,\widetilde{\psi}^{(1)}(\bfp - \bfp')\;.
\ee
We stress that to obtain the expression of $\widetilde{\psi}^{(2)}$ in terms of
the first--order Zel'dovich potential $\widetilde{\psi}^{(1)}$ it is completely
unnecessary, for practical uses, to specify the form of the divergence--free
vector $\bfR^{(2)}$, for any realistic initial conditions. The kernel
$\kappa^{(2)}(\bfp_1,\bfp_2)$ describes in the Lagrangian Fourier space the
effects of the non--linear dynamics.

Similar considerations and calculations may be easily extended to the
third--order solutions. We give here the explicit expressions of the Fourier
components of the Lagrangian potentials $\psi^{(3)}_a$ and $\psi^{(3)}_b$,
originating the longitudinal motion, and of the vector potential $\bfA$,
describing the transversal motion.

To show how the calculations progress, it is more simple to start with the
derivation of the Fourier component $\widetilde{\psi}^{(3)}_b$. From the
expression (44),
\be
\widetilde{\psi}^{(3)}_b(\bfp)
=-\f{p_{\al}}{2\,p^2}\int \f{d\bfp_1 d\bfp_2}{(2\pi)^6}
\Big[(2\pi)^3 \delta_D(\bfp_1+\bfp_2 - \bfp)\Big]\,
\Big(p^2_2\,p_{1\al}-\bfp_1\cdot\bfp_2\,p_{2\al}\Big)
\,\,\widetilde{\psi}^{(1)}(\bfp_1)\,\widetilde{\psi}^{(2)}(\bfp_2)\;.
\ee
Inserting the solution (66), we eventually obtain, in terms of the
Zel'dovich potential, the expression
\be
\widetilde{\psi}^{(3)}_b(\bfp)=
\int \f{d\bfp_1 d\bfp_2}{(2\pi)^6}
\,\,\kappa^{(3)}_b(\bfp_1,\bfp_2; \bfp)\,\,
\widetilde{\psi}^{(1)}(\bfp_1)\,\widetilde{\psi}^{(1)}(\bfp_2 - \bfp_1)
\,\widetilde{\psi}^{(1)}(\bfp - \bfp_2)\;,
\ee
where the kernel $\kappa^{(3)}_b$ has been introduced:
\be
\kappa^{(3)}_b(\bfp_1,\bfp_2;\bfp)\equiv
\f{1}{2}\,\kappa^{(2)}(\bfp_1,\bfp_2)
\left[1-\left(\f{\bfp\cdot\bfp_2}{p\,p_2}\right)^2\right]\;.
\ee
In a similar fashion, the Fourier component $\widetilde{\psi}^{(3)}_a$
may be derived:
\be
\widetilde{\psi}^{(3)}_a(\bfp)= -\f{1}{p^2}
\int \f{d\bfp_1 d\bfp_2}{(2\pi)^6}
\,\,\kappa^{(3)}_a(\bfp_1,\bfp_2; \bfp)\,\,
\widetilde{\psi}^{(1)}(\bfp_1)\,\widetilde{\psi}^{(1)}(\bfp_2)
\,\widetilde{\psi}^{(1)}(\bfp - \bfp_1 - \bfp_2)\;,
\ee
where, again, we have defined a kernel $\kappa^{(3)}_a$ according to the
definition
\be
\kappa^{(3)}_a(\bfp_1,\bfp_2; \bfp) \equiv
\f{1}{6}\,\Big[\epsilon_{\al\gamma\de}\,p_{\al}\,p_{1\gamma}\,p_{2\de}\Big]\,
\Big[\epsilon_{\beta\eta\sigma}\,
(p_{\beta}-p_{1\beta}-p_{2\beta})\,p_{1\eta}\,p_{2\sigma}\Big]\;.
\ee
The algebraic relation $S^{(1)C}_{\al\beta} =
\f{1}{2}\epsilon_{\al\gamma\de}\,\epsilon_{\beta\eta\sigma}\,
S^{(1)}_{\gamma\eta}\,S^{(1)}_{\delta\sigma}$ has to be used. Finally,
the vortical component $\widetilde{T}_{\al}$ along the
$\hat{\al}$--direction is given by the integral
\be
\widetilde{T}_{\al}(\bfp)= i\int \f{d\bfp_1 d\bfp_2}{(2\pi)^6}
\,\,\iota^{(3)}_{\al}(\bfp_1,\bfp_2; \bfp)\,\,
\widetilde{\psi}^{(1)}(\bfp_1)\,\widetilde{\psi}^{(1)}(\bfp_2 - \bfp_1)
\,\widetilde{\psi}^{(1)}(\bfp - \bfp_2)\;,
\ee
where the kernel $\iota^{(3)}_{\al}$, which indeed depends upon the
direction, is explicitly given by
\be
\iota^{(3)}_{\al}(\bfp_1,\bfp_2;\bfp)\equiv
\f{1}{2}\,\kappa^{(2)}(\bfp_1,\bfp_2)\left(1-\f{\bfp\cdot\bfp_2}{p_2^2}\right)
\left(2\bfp_2-\bfp \right)_{\al}\;.
\ee
Specifically, the vortical vector $\widetilde{\bfT}(\bfp)$
has only two independent components, in that it satisfies the condition
of transversality, $\bfp\cdot\widetilde{\bfT}=0$. Therefore it follows
that the vector potential $\bfA$ is fully specified by only two of its
three components: without loss of generality we can decide that $A_z\equiv 0$.
The significant Fourier components $\widetilde{A}_x$ and $\widetilde{A}_y$
are thus given by the relations:
\be
p_z \,\widetilde{A}_y(\bfp) = i\,\widetilde{T}_x(\bfp)\;,
\ee
and
\be
p_z\,\widetilde{A}_x(\bfp) = i\,\widetilde{T}_y(\bfp)\;.
\ee
These last two equations complete the analysis of the Lagrangian
motion up to the third--order perturbative approximation: specifically,
the third--order dynamics is fully
described in terms of the tetra--potential
$\Psi\equiv (\psi^{(3)}_a, \psi^{(3)}_b, A_1, A_2) $. The complications
due to the non--linear evolution are summarized in the five kernels
$\kappa^{(2)}$, $\kappa^{(3)}_a$, $\kappa^{(3)}_b$ and $\iota^{(3)}_{\al}$, the
two last ones depending on the chosen direction.

\bc
\section*{Appendix D}
\ec
In this last appendix, and as an example of how our formalism works, we want
to show that the general irrotationality condition in Lagrangian space
\be
\eps_{\al\beta\gamma}\,\dot{S}_{\beta\sigma}
\left[ (1+\nabla\cdot\bfS)\,\de_{\gamma\sigma} - S_{\gamma\sigma}
+ S^C_{\gamma\sigma} \right] = 0\;,
\ee
leads to the eq.(46) for the third--order transverse component
$\bfT$ of the displacement $\bfS$. From the ansatz (45) one
immediately obtains
\be
\dot{S}_{\beta\sigma}=
\dot{D}(\tau)\,S^{(1)}_{\beta\sigma}+
\dot{E}(\tau)\,S^{(2)}_{\beta\sigma}+
\dot{F}_c(\tau)\,T_{\beta\sigma}\;,
\ee
(the $F_a$--mode and $F_b$--mode do not enter in the vortical couplings) and,
from eq.(61), we get
$$
0 = - \dot{F}_c\,\eps_{\al\beta\gamma}\, T_{\gamma\beta}\; +
$$
\be
+\;\eps_{\al\beta\gamma}\left[\dot{D}S^{(1)}_{\beta\sigma}
+\dot{E}S^{(2)}_{\beta\sigma} \right]
\left[D\mu_1(1)\,\de_{\gamma\sigma}
+ E\mu_1(2)\,\de_{\gamma\sigma}
-D S^{(1)}_{\gamma\sigma} - E S^{(2)}_{\gamma\sigma} +
D^2 S^{(1)C}_{\gamma\sigma}\right]\,,
\ee
i.e.
\be
\eps_{\al\beta\gamma}\left\{\dot{D}
D^2 S^{(1)}_{\beta\sigma}\,S^{(1)C}_{\gamma\sigma}
+
\dot{D}ES^{(1)}_{\beta\sigma}
\left[\mu_1(2)\,\de_{\gamma\sigma} -S^{(2)}_{\gamma\sigma} \right]
+ D\dot{E}S^{(2)}_{\beta\sigma}
\left[\mu_1(1)\,\de_{\gamma\sigma} -S^{(1)}_{\gamma\sigma} \right]
-\dot{F}_c\, T_{\gamma\beta}
\right\} = 0\,,
\ee
where for brevity we wrote $\mu_1(n)\equiv \mu_1(\bfS^{(n)})$.
Recalling that $\eps_{\al\beta\gamma}S^{(1)}_{\beta\gamma}=0$,
$\eps_{\al\beta\gamma}S^{(2)}_{\beta\gamma}=0$ and
$\eps_{\al\beta\gamma}S^{(1)}_{\beta\sigma}S^{(1)C}_{\gamma\sigma}=0$,
we finally obtain
\be
\eps_{\al\beta\gamma}\left[(D\dot{E}-\dot{D}E)S^{(1)}_{\beta\sigma}
S^{(2)}_{\sigma\gamma} - \dot{F}_c\, T_{\gamma\beta}\right] = 0\;,
\ee
which coincides with the relation (46) in the main text.
\end{document}